\documentclass[aps,twocolumn,nopacs,floats,superscriptaddress]{revtex4-1}

\usepackage[colorlinks=true, citecolor=blue, linkcolor=blue, urlcolor=blue]{hyperref}
\usepackage{graphicx,dcolumn,longtable,epsfig}
\usepackage[usenames]{color}
\usepackage{amssymb}
\usepackage{amsmath}
\usepackage{bm}
\usepackage{footnote}
\usepackage{float}
\usepackage{subfigure}
\usepackage{color}

\usepackage{ulem}
\usepackage[T1]{fontenc}

\newcommand{\be}{\begin{equation}}
\newcommand{\ee}{\end{equation}}

\newcommand{\ket}[1]{\vert #1 \rangle}
\newcommand{\bra}[1]{\langle #1 \vert}
\newcommand{\mel}[3]{\langle #1 \vert #2 \vert #3 \rangle}
\newcommand{\ov}[2]{\langle #1 \vert #2 \rangle}

\def\bea{\begin{eqnarray}}
\def\eea{\end{eqnarray}}

\begin{document}
\title{Bond Bipolarons:  Sign-free Monte Carlo Approach}

\author{Chao Zhang}
\affiliation{State Key Laboratory of Precision Spectroscopy, East China Normal University, Shanghai 200062, China}
\author{Nikolay V. Prokof'ev}
\affiliation{Department of Physics, University of Massachusetts, Amherst, Massachusetts 01003, USA}
\author{Boris V. Svistunov}
\affiliation{Department of Physics, University of Massachusetts, Amherst, Massachusetts 01003, USA}
\affiliation{National Research ``Center Kurchatov Institute,'' 123182 Moscow, Russia}
\affiliation{Wilczek Quantum Center, School of Physics and Astronomy and T. D. Lee Institute, Shanghai Jiao Tong University, Shanghai 200240, China}

\begin{abstract}
Polarons originating from phonon displacement modulated hopping have relatively light masses and, thus, are
of significant current interest as candidates for bipolaron mechanism of high-temperature superconductivity
[Phys. Rev. Lett. {\bf 121}, 247001 (2018)]. We observe that the bond model, when the dominant coupling comes from
atomic vibrations on lattice bonds, can be solved by efficient sign-free Monte Carlo methods based on the path-integral formulation of the particle sector in combination with either the (real-space) diagrammatic or Fock-path-integral representation of the phonon sector. We introduce the corresponding algorithms and provide illustrative results for
bipolarons in two dimensions. The results suggest  that the route towards high-temperature superconductivity (if any) in the multiparametric space of the model lies between the Scylla of large size of moderately
light bipolarons and Charybdis of large mass of compact bipolarons. As a result, on-site repulsion is helping
$s$-wave superconductivity in sharp contrast with existing expectations.
\end{abstract}

\maketitle

 {\it Introduction.} Broad interest in polarons---stable quasiparticles emerging as a result of strong renormalization
of properties of a bare particle due to interaction with this or that environment---comes from their ubiquitous presence across all fields of physics.
There exist electron-phonon polarons~\cite{Landau33,Frohlich50,Feynman55, Schultz1959, Holstein59, Alexandrov99, Holstein2000, Mona2010},
spin-polarons~\cite{BR70,Nagaev74, Mott06}, Fermi-polarons~\cite{Lobo06, Bulgac07},
protons in neutron rich matter \cite{protons}, etc.
Coupling to the environment may also lead to formation of bound particle states called bipolarons.
Their studies are motivated by the possibility of identifying a possible mechanism for high-temperature
superconductivity when a gas/liquid of bound pairs of fermions undergoes a superfluid (SF) transition at temperature
\be
T_c \sim {n^{2/d}\over m_{\rm *} } \, ,
\label{Tc}
\ee
where $n$ is the bipolaron density in a $d\ge 2$-dimensional system and $m_{\rm *}$ is the bipolaron effective mass.
The estimate for $T_c$ is very robust against repulsive interactions between bosons: in $d=3$ it barely changes between
the gas and liquid densities \cite{Ceperley,Landau,tc3da,tc3daa,tc3db}, and in $d=2$ the dependence on interactions
is logarithmically weak \cite{tc2da,tc2daa,tc2db,tc3db}. Since the SF transition temperature increases with density, the
highest value of $T_c$ for this mechanism is obtained when the inter-particle distance is of the order of the pair
size, or $n R_*^d \sim 1$, where $R_*$ is the bipolaron radius.
By increasing the fermion density further one goes through the so-called BEC-to-BCS crossover corresponding to a radical change of the microscopic mechanism behind the SF transition---the (quasi) Bose-Einstein condensation (BEC) of spatially separated pairs gets  gradually replaced by the Bardeen-Cooper-Schreiffer (BCS) pairing in momentum space---and $T_c$
starts decreasing exponentially \cite{Leggett,BCSBEC}. Thus, the maximum value
of transition temperature can be expressed though the bipolaron parameters as
\be
T_c \sim {1\over m_{\rm *}R_*^2 }\, .
\label{maxTc}
\ee

When the dominant mechanism of electron-phonon coupling is of the density-displacement type, as in the Holstein model
\cite{Holstein59}, large values of $T_c$ cannot be reached \cite{Chakraverty, PhysRevLett.84.3153, PhysRevB.69.245111}.
The reason is exponentially large effective bipolaron masses originating from small phonon overlap integrals
for realistic values of the adiabatic parameter
\[
\gamma\, =\, {\omega_{\rm ph}\over W} \, \ll\,  1 ,
\]
where  $\omega_{\rm ph}$ is the phonon frequency and $W$ is the particle bandwidth; $W \approx 4dt$ for tight-binding dispersion relation with hopping amplitude $t$ between the nearest-neighbor sites on a simple $d$-dimensional cubic lattice. The corresponding bipolarons are also very sensitive
to the on-site Hubbard repulsion $U$. Much stronger electron-phonon interaction (EPI) is required for their
formation when $U \sim W$, leading to an additional exponential increase of the effective mass
\cite{PhysRevLett.84.3153}.

The prospects for obtaining high $T_c$ appear to be far better when strong EPI originates from
hopping dependence on atomic displacements by one of the two mechanisms. In mechanism A, tunneling
is enhanced (suppressed) when the distance between the sites is reduced (increased)~\cite{PhysRevLett.25.919, PhysRevB.5.932}.
In mechanism B, tunneling is modulated by displacements of atoms in the barrier region between the sites~\cite{KK}.
Previous studies \cite{Mona2010,us21,Sous21} found that these bond polarons remain relatively
light when entering the strong coupling regime. However, properties of the corresponding
bipolarons remain unexplored: the only available study \cite{Sous2018} considered the
$d=1$ case in the antiadiabatic limit $\omega_{\rm ph} = 3t$ for mechanism A.
For realistic predictions of high $T_c$ one needs to look at higher dimensional systems in the
adiabatic limit $\gamma \ll 1$ and account for large repulsive
potential between the electrons. This poses a significant computational challenge for unbiased methods
based on exact diagonalization \cite{PhysRevLett.84.3153,Bonca2001} or controlled truncation of the phonon
Hilbert space \cite{Carbone2021} because bipolaron states in $d>1$ are extended and the number of excited
phonon modes quickly increases with $1/\gamma$.

In this Letter, we show that path-integral representation for particles offers a unique opportunity for
conducting comprehensive studies of $n$-particle polaronic states when EPI is based on mechanism B. The simplest
Hamiltonian can be formulated as $\hat{H} = \hat{H}_{e}+\hat{H}_{\rm ph}+\hat{H}_{\rm int}$ with
 \begin{equation}
\hat{H}_{e}  =  - t \! \! \sum_{<ij>,\sigma}  \! \!   (\hat{c}_{j \sigma}^{\dagger} \hat{c}_{i\sigma}^{\;} +{\rm H.c.}) +  \!   \sum_{ij \sigma \sigma'}
\! V_{\sigma \sigma'}(i,j) \hat{n}_{i \sigma} \hat{n}_{j \sigma'} ,
\label{He}
\end{equation}
 \begin{equation}
\hat{H}_{\rm ph} = \omega_{\rm ph} \sum_{<ij>} \, (\hat{b}_{<ij>}^{\dagger} \hat{b}_{<ij>}^{\;} +1/2) \, ,
\label{Hph}
\end{equation}
 \begin{equation}
\hat{H}_{\rm int} = -g \! \sum_{<ij>,\sigma} \, (\hat{c}_{j \sigma}^{\dagger} \hat{c}_{i\sigma}^{\;} +{\rm H.c.})(\hat{b}_{<ij>}^{\dagger} + \hat{b}_{<ij>}^{\;}) ,
\label{Hint}
\end{equation}
where $V_{\sigma \sigma'}(i,j)$ is the potential of interelectron interaction and $g$ is the strength of EPI.
We use standard notation for creation and annihilation operators for electrons (on site $i$ with spin $\sigma$)
and optical phonons (on bonds $<ij>$). The model can be further generalized to deal with several dispersive
phonon modes and study effects of phonon dynamics and polarization. Here we focus on describing the numerical method 
and present illustrative results for bipolaron states in a two-dimensional system. These results
indicate that the search for high-$T_c$ regimes requires comprehensive exploration of the model parameter space
because compact states can easily end up to be heavy while light effective masses come at the price of larger
pair radius, see Eq.~(\ref{maxTc}). One counter-intuitive effect is that the $s$-wave transition temperature
may increase with the on-site repulsion $U$ because less compact bipolarons are significantly lighter.


{\it Basic relations.} For a generic few-body system, the mathematical object containing all information
about ita ground-state properties and potentially allowing {\it sign-free} Monte Carlo simulations can be
formulated as
\be
O_{ba}(\tau )  =  \mel{\, b\; }{e^{- (\tau /2) \hat{H} } \, \hat{O} \, e^{- (\tau /2) \hat{H} }}{\, a\, } \, .
\label{O}
\ee
Here $\hat{O}$ is a certain observable, $\hat{H}$ is the system's Hamiltonian, $\tau$ is an appropriately large
imaginary-time interval, $\ket{\, a\, }$ and $\ket{\, b\, }$ are any two states having finite overlap with the
ground state, $\ket{{\rm g}}$.  In the asymptotic limit of $\tau \to \infty $ the answer is dominated
by the projection onto the ground state when $ O_{ba}(\tau ) $ is given by the product of the
ground-state expectation value of $\hat{O}$ and the universal (for all observables) propagator ${\cal G}_{ba}(\tau)$:
\be
O_{ba}(\tau )  \; \mathop  {\longrightarrow} \limits_{\tau \to \infty}  \;
  \mel{\, {\rm g}\, }{\, \hat{O} \, }{\, {\rm g}\, } \,  \mel{\, b\, }{e^{- \tau \hat{H}}}{\, a\, }
  \equiv \bar{O}_{\rm g} \, {\cal G}_{ba}(\tau) \, .
\label{O1}
\ee
If the states  $\ket{\, a\, }$ and $\ket{\, b\, }$ are free of phonons, ${\cal G}_{ba}$ is the standard
$n$-particle Green's function.
Since ${\cal G}_{ba}(\tau) \equiv  I_{ba}(\tau) $, where $\hat{I}$ is the identity operator,
the expectation value of $\hat{O}$ in the ground state can be represented as
\be
   \bar{O}_{\rm g} \, = \, \frac{O_{ba} (\tau ) } { I_{ba} (\tau )  }
\,    \equiv \, \frac{\sum_{ab} W_{ab} \, O_{ba} (\tau ) } { \sum_{ab} W_{ab} \, I_{ba} (\tau )  }  \, .
\label{O_MC}
\ee
This is standard for MC methods setup when the stochastic sampling process is designed to sample
the propagators ${\cal G}_{ba}$ while matrix elements of physical properties are taken care of
by Monte Carlo (MC) estimators. The extended version of the relation---with sums over any subsets of states $\ket{\, a\, }$ and $\ket{\, b\, }$
with arbitrary weights $W_{ab}$---adds flexibility and efficiency in designing
updating schemes.

According to Eqs.~(\ref{O}) and (\ref{O1}), the imaginary-time dependence of ${\cal G}_{ba}$
also contains direct information about the ground-state energy, $E_{\rm g}$:
\be
{\cal G}_{ba}(\tau )  \; \mathop  {\longrightarrow} \limits_{\tau \to \infty} \;
\ov{\, b\, }{\, {\rm g}\, }  \ov{\, {\rm g}\, } {\, a\, } \, e^{- \tau E_{\rm g}} .
\label{F_energy}
\ee
Moreover, by selecting the state $\ket{\, b\, }$ or state  $\ket{\, a\, }$, or both,
to belong to a particular---non-ground-state---symmetry sector, we can employ (\ref{F_energy})
to determine energy of the ground state in a given sector. This way ${\cal G}_{ba}(\tau)$ can be 
used to obtain the quasiparticle dispersion relation, and, in particular, its effective mass.

A direct procedure of extracting $m_{\rm *}$ from ${\cal G}_{ba}(\tau)$ is based on the coordinate representation
when for each of the states $\ket{\, b\, }$ and $\ket{\, a\, }$ we introduce the notion of the ``center-of-mass''
positions, ${\bf R}_b$ and ${\bf R}_a$, respectively, and consider the relative-coordinate dependence
of the propagator ${\cal G}_{ba} (\tau, \, {\bf R} )$, where ${\bf R}={\bf R}_b - {\bf R}_a$.
In the limit of long time and large distance, $( \tau, {\bf R}^2 )\to  \infty$,
this dependence takes on the characteristic form of a single free particle propagator with effective mass $m_*$:
\be
{\cal G}_{ba} (\tau , {\bf R}) \,  \to  \,  {A_{ba}   e^{-E_{\rm g} \tau}  \over \tau^{d/2}} \, e^{-{m_* {\bf R}^2 \over 2 \tau}} .
\label{Gauss}
\ee
Apart from the coefficient $A_{ba}$, the r.h.s. of (\ref{Gauss}) is insensitive to the particular
choice of states $\ket{\, a\, }$ and $\ket{\, b\, }$, allowing one to average over them as in Eq.~(\ref{O_MC}).
In the asymptotic limit, the difference between the lattice and continuous space disappears, and $m_*$
is directly related to the mean square fluctuations of the relative coordinate:
\be
\overline{{\bf R}^2} (\tau )  = \frac{ \sum_{ab} W_{ab} \, {\cal G}_{ba} (\tau, {\bf R}) \, {\bf R}^2 }
                                     { \sum_{ab} W_{ab} \, {\cal G}_{ba} (\tau, {\bf R})           }
\;  \mathop  {\longrightarrow} \limits_{\tau \to \infty} \; \frac{d}{m_*}\tau \, .
\label{R2}
\ee
Starting with zero at $\tau = 0$, the $\overline{{\bf R}^2} (\tau)$ function ultimately saturates to a
straight line at long $\tau$ leading to an accurate estimate of the effective mass from its slope.

%
%

{\it Perturbative expansion.} Our approach to MC sampling is based on the general scheme
proposed in Ref.~\cite{worm} and rendered here. Let $\hat{H}_0$ and $\hat{V}$ be the diagonal and off-diagonal parts of the
Hamiltonian with respect to some basis ${\cal B} = \{ \ket{\alpha} \} $
(out choice is site Fock states for particles and bond Fock states for phonons):
$\hat{H}_0 \ket{\alpha} = E_{\alpha}  \ket{\alpha}$,  $\bra{\alpha } \hat{V} \ket{\alpha} =0$.
Next, we decompose $\hat{V}$ into elementary non-vanishing terms,
$\hat{V} = \sum_{\alpha \beta} V_{\beta \alpha} \ket{\beta} \bra{\alpha }$.
In the chosen representation, there are three types of the elementary terms
each corresponding to the particle hopping along a specific bond
in a specific direction: (i) bare hopping
(ii) hopping assisted by the phonon creation, and (iii) hopping assisted by the phonon annihilation.
For a sign-free MC method---apart from, possibly, the negative signs (or phases) originating from components
of the vectors $\ket{\, a\, }$ and $\ket{\, b\, }$ in the basis ${\cal B}$---all matrix
elements $V_{\beta \alpha}$ need to be non-negative real numbers.
This requirement is satisfied with our choice of ${\cal B}$ for model (\ref{He})-(\ref{Hint}).

Using standard interaction representation for the evolution operator in the imaginary time domain,
$e^{-\tau \hat{H}}  = e^{-\tau \hat{H}_0} \hat{\sigma } (\tau) $,
and expanding $\hat{\sigma } (\tau)$ into Taylor series we arrive at \cite{worm}:
\begin{eqnarray}
&  & \sigma_{\beta \alpha}(\tau)  =
\delta_{\alpha \beta} \, +  \int_0^{\tau}
d \tau_1 \: V_{\beta \alpha} \, e^{\tau_1 E_{\beta \alpha}} \nonumber
\\
& + & \sum_{\gamma_1} \!  \int_{0}^{\tau}  \! \! \!  d \tau_2 \! \!
                       \int_0^{\tau_2}  \! \! \!  d \tau_1
V_{\beta \gamma_1 }  e^{\tau_2 E_{\beta    \gamma_1}}
V_{\gamma_1 \alpha} e^{\tau_1 E_{\gamma_1 \alpha}  } + \ldots  ,  \qquad
\label{sigma}
\end{eqnarray}
where $E_{\beta \alpha} = E_{\beta} - E_{\alpha}$.
The MC scheme is based on the statistical interpretation of the r.h.s. of (\ref{sigma}) viewed as an
average over an ensemble of graphs representing strings of the hopping transitions, or ``kinks'',
$V_{\gamma_{i+1} \gamma_{i}}$.
A string is characterized by the number and types of kinks, as well as by their space-time positions.
Graphs are sampled according to their non-negative weights given by the values of the corresponding
integrands in Eq.~(\ref{sigma}).

{\it Sampling protocol and estimators.}
Our MC scheme includes $\tau$ into the configuration space of graphs and allows us, on the one hand,
to sample the $\tau$-dependence of ${\cal G}_{ba}(\tau)$ and $\overline{{\bf R}^2} (\tau)$ and then use Eqs.~(\ref{F_energy}) and (\ref{R2}) to estimate the ground-state energy and the effective mass. On the other hand,
this setup dramatically simplifies the minimalistic set of updates that deal exclusively with the last
(in the $\tau$-domain) kink in the string. To be specific, we stochastically (i) add and remove the last bare-hopping
kink using a pair of complementary updates \cite{worm}, (ii) switch between the three types of the hopping terms, and
(iii) sample the length of the last time interval separating the last kink from the state $\bra{\, b\,}$.
This scheme is ergodic and produces states $\bra{\, b\,}$ that admit any allowed configuration of excited phonon modes.
Clearly, one can add additional updates dealing with other than last kinks for better decorrelation of the sting configuration, see \cite{worm}.

The proposed setup provides access to all the relevant system's properties, including the same- and
different-time correlation functions. They are measured when the value of $\tau$ is large enough and
the asymptotic limit (\ref{O1}) is reached; its maximal value $\tau_{\rm max}$ is an important
parameter controlling the accuracy of the projection onto the ground state.
Monte Carlo estimators for observables are based on Eqs.~(\ref{O}) and (\ref{O_MC}) and their straightforward
generalization for the different-time correlators. In accordance with the general theory (see, e.g., Refs.~\cite{worm})
estimators do not involve additional computational costs (i.e. modifications of the configuration
space and the sampling protocol) when the corresponding operators are either (i) diagonal in the basis
${\cal B}$ or (ii) expressed in terms of elementary $\hat{V}$-terms. The estimator for observable of type (i)
is simply the eigenvalue of this observable for the state $\ket{\gamma} \in {\cal B}$ created by the
string of kinks at a certain moment of imaginary time, $\tau_*$, close to $\tau/2$.
The estimator for observable of type (ii) is minus the total number of the corresponding
$\hat{V}$-kinks within a certain time interval divided by the duration of this interval.
It can be arbitrarily long provided its boundaries are appropriately far from the string end points
$\tau=0$ and $\tau \sim \tau_{\rm max}$.

{\it Diagrammatics for the phonon sector.} In the considered model, phonons do not interact with each other.
This allows one to treat them diagrammatically, while keeping the path-integral representation for the particle sector
only. The benefit of the diagrammatic representation for phonons (compared to their path-integral treatment
described above) is the flexibility of dealing with dispersive modes.
Here the phonon dispersion is accounted for by simply modifying the zero-temperature
(to comply with the diagrammatic rule that all averages be specified in terms of the vacuum state)
phonon propagators, $D_s({\bf r}_1,\tau_1; {\bf r}_2,\tau_2) \equiv D_s ({\bf r}_1-{\bf r}_2,\tau_1,\tau_2) $,
with $s=1, \dots, d$ enumerating directions of lattice bonds.
For sign-free formulation, $D_s$ should be non-negative.
One can still access (i) the particle configurations, (ii)
the distribution of phonon numbers, and the correlations between (i) and (ii).
The path-integral treatment of dispersive phonons is also possible,
and comes at a price of dealing with yet another family of kinks originating from phonon hopping. The benefit of
this approach is detailed information about spatial configurations of excited phonon modes.

The mathematical structure of the perturbative expansion with diagrammatic treatment of phonons is similar to
the series (\ref{sigma}) provided the Greek subscripts are used exclusively for the particle states.
Accordingly, $E_\alpha$ now refer to energies of the particle subsystem alone, while each kink in (\ref{sigma})
representing the phonon-assisted hopping event is associated with one of the two end-points of the vacuum phonon
propagator $D_s({\bf r}_1,\tau_1; {\bf r}_2,\tau_2)$. The other end belongs either to states
$\ket{\, a\, }$ or $\bra{\, b\, }$ or another phonon-assisted hopping event along the bond with the same direction $s$.

\begin{figure}[t!]
\centering
\includegraphics[width=0.23\textwidth]{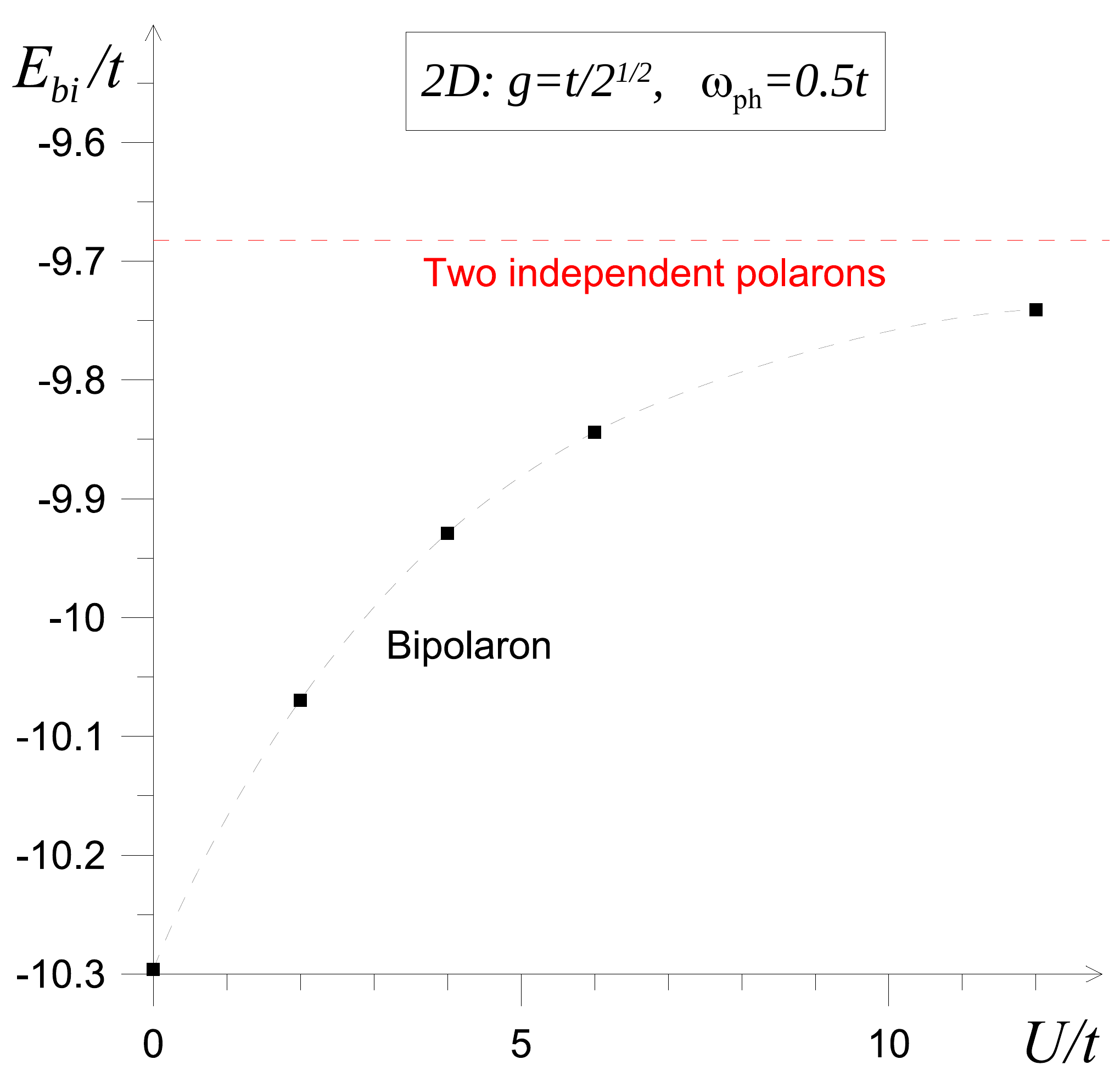}
\includegraphics[width=0.23\textwidth]{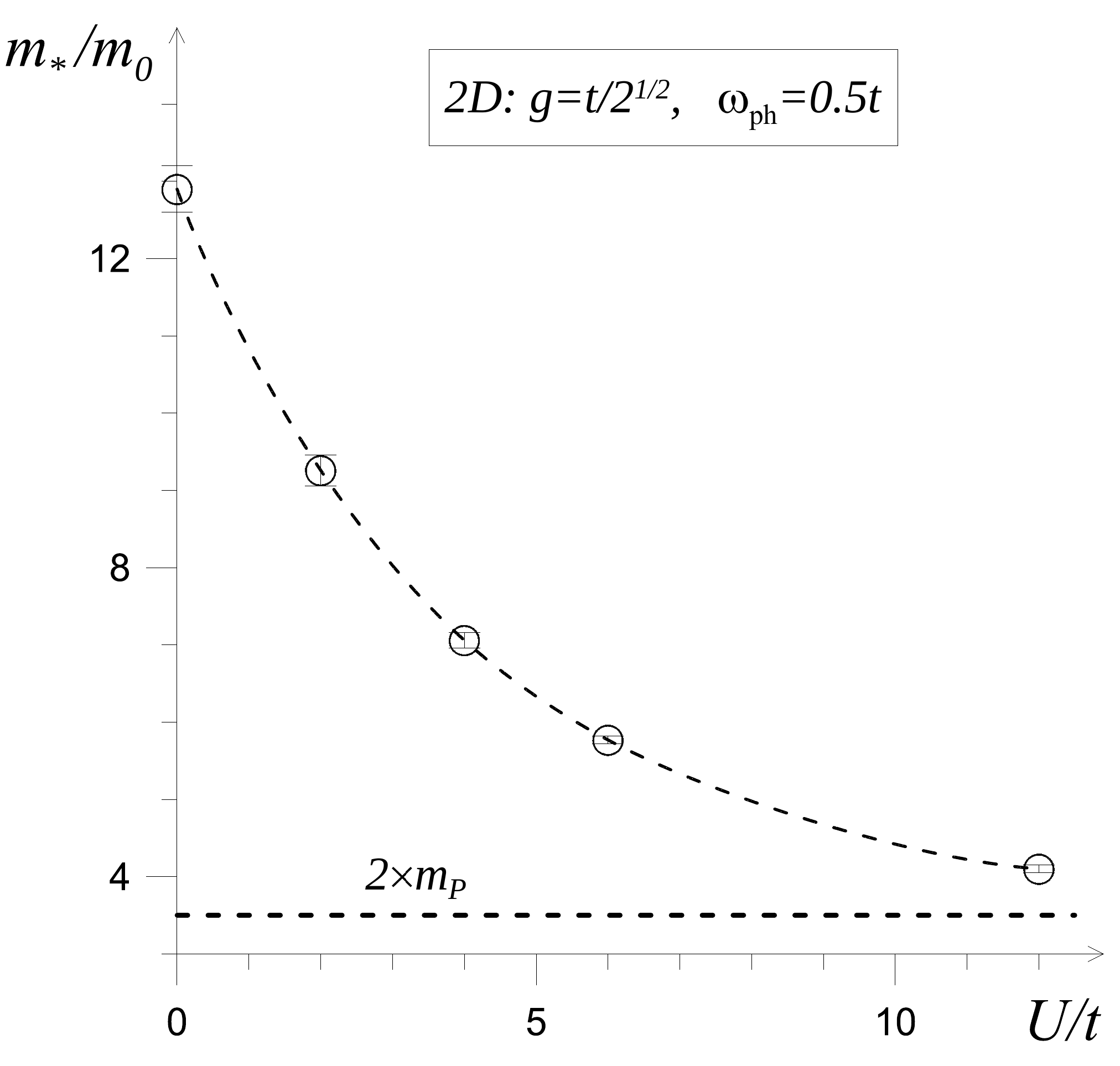}
\caption{Bipolaron energies (left) and effective masses (right) as functions of on-site
repulsion $U$ for coupling constant $g=t/\sqrt{2}$ and adiabatic parameter $\gamma=1/16$.
Error bars for $E_{bi}$ are much smaller than symbol size. Dashed lines in all plots are
used to guide an eye.
}
\label{fig1}
\end{figure}

{\it Illustrative results}.
For estimates of $T_c$, apart from the bi-polaron energy and effective mass,
we determine $R_*^2$ from the probability distribution $P(R_{12})$ of finding particles
at distance $R_{12}$ from their center of mass position
\begin{equation}
R_*^2 \, = \, \sum_{R_{12}} R_{12}^2 \, P(R_{12}) \,.
\label{R2}
\end{equation}
Figure~\ref{fig1} shows the dependence of the ground-state energy and the effective mass on the strength of the on-site
repulsion for coupling constant $g=t/\sqrt{2}$ and the adiabatic parameter $\gamma=1/16$. The ground-state energies are found
to be smaller than the energies of two polarons for any realistic value of $U$, indicating that we are dealing with bound states
(bipolarons). At $U=0$, the bipolaron effective mass is more than three times larger than its asymptotic limit $2m_p$,
where $m_p$ is the mass of a single polaron. With increasing $U$, the effective mass drops significantly, and at  $U\gtrsim 10$ it approaches the limiting value of $2m_p$.

\begin{figure}[t]
\centering
\includegraphics[width=0.23\textwidth]{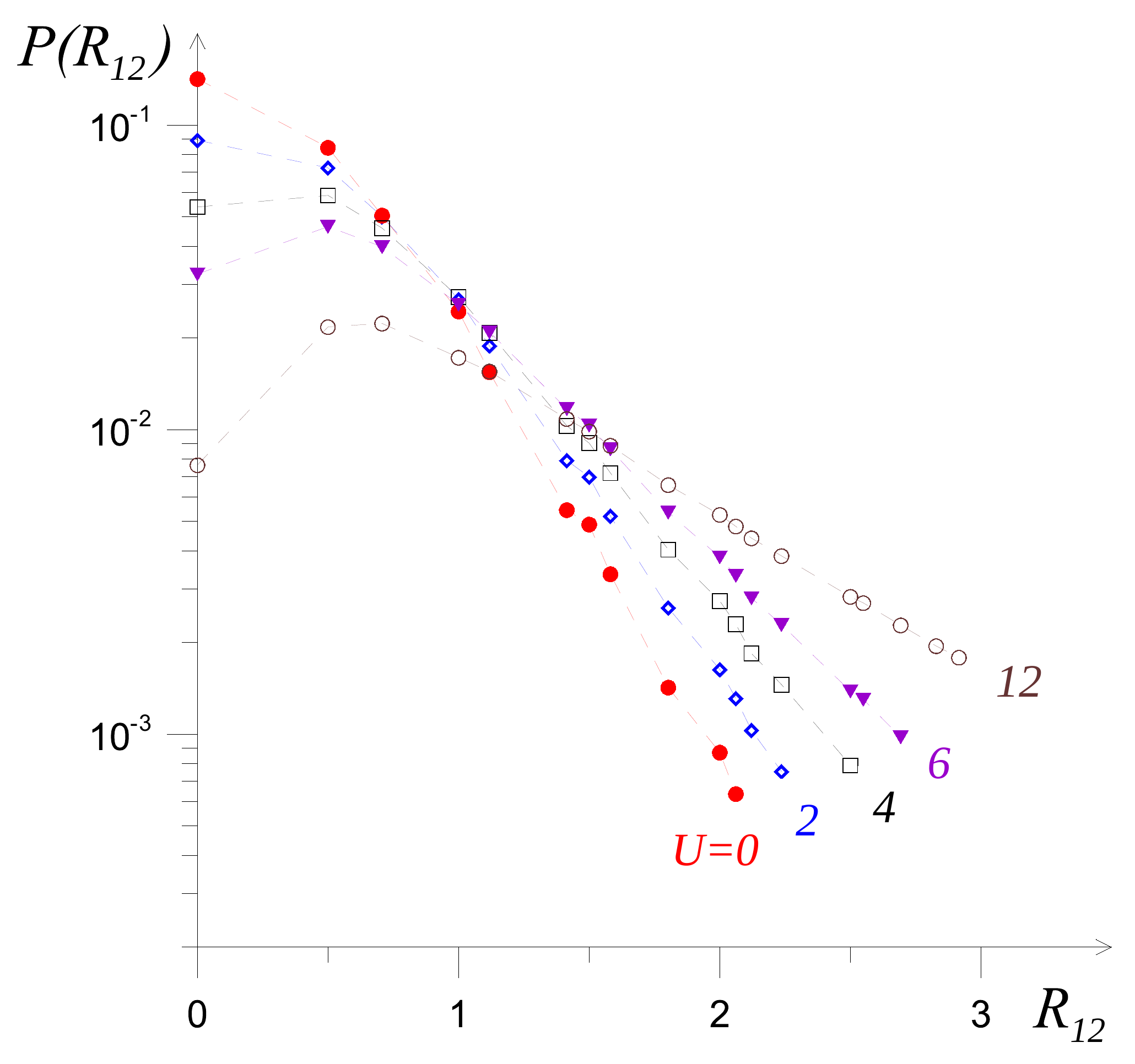}
\includegraphics[width=0.23\textwidth]{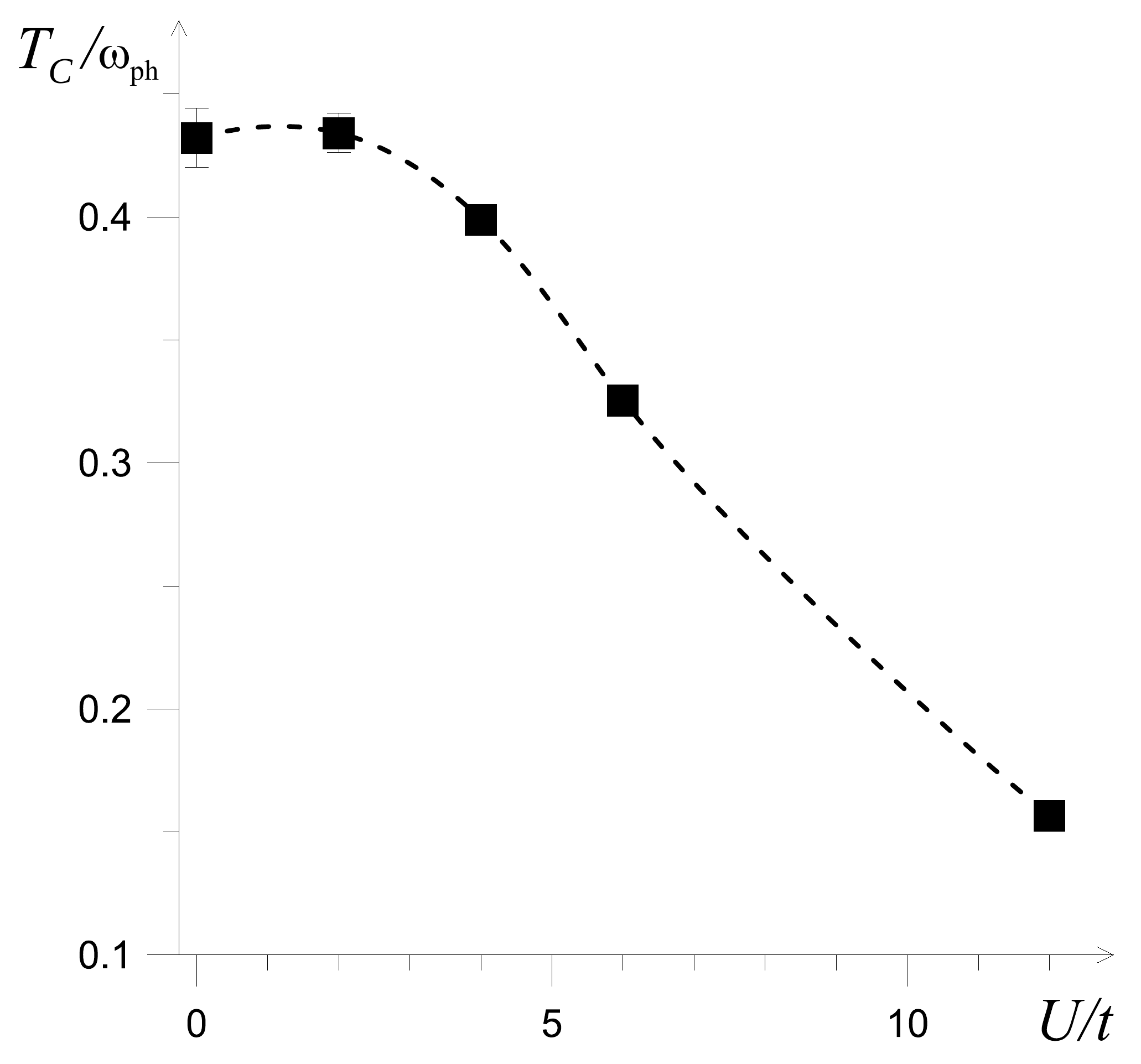}
\caption{Radial profiles for bipolaron states (left) and estimates of $T_c$ (right)
based on Eq.~(\ref{maxTc}) for the same set of parameters as in Fig.~\ref{fig1}.
}
\label{fig2}
\end{figure}

In contrast to naive expectations that a significant drop in the effective mass should result in a substantial
increase of $T_c$, the trend revealed in Fig.~\ref{fig2} is quite different. The estimate for the critical
temperature remains relatively flat up to $U\sim 2$ and then starts to decrease, dropping by almost a factor of 3 at $U \sim 10$. This is because the decrease of $m_*$ is accompanied by a rapid increase of $R_*$, so that the product $m_*R_*^2$
does not show a pronounced maximum at any finite $U$.
\begin{figure}[t]
\centering
\vspace*{0.1cm}
\includegraphics[width=0.23\textwidth]{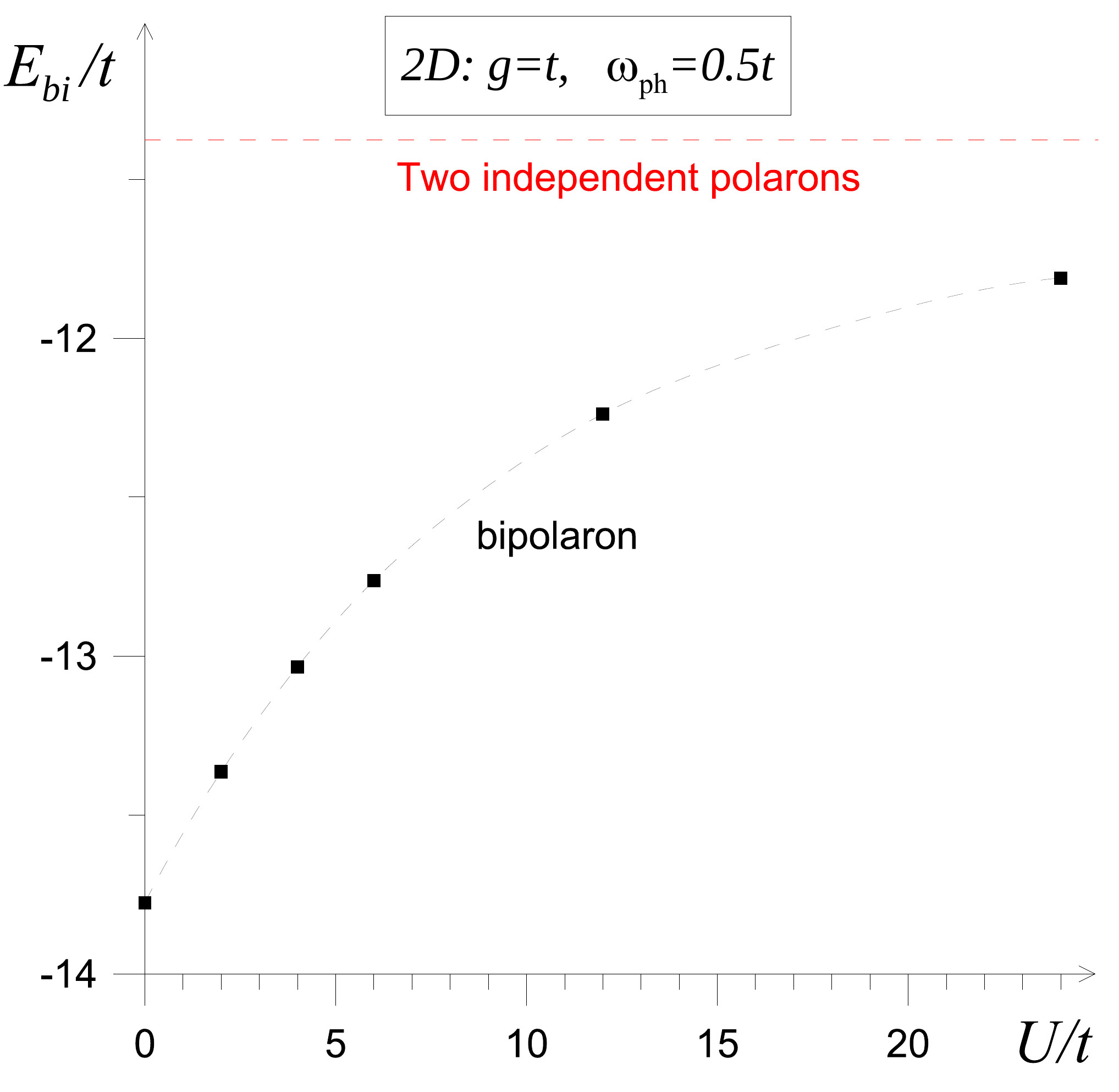}
\includegraphics[width=0.23\textwidth]{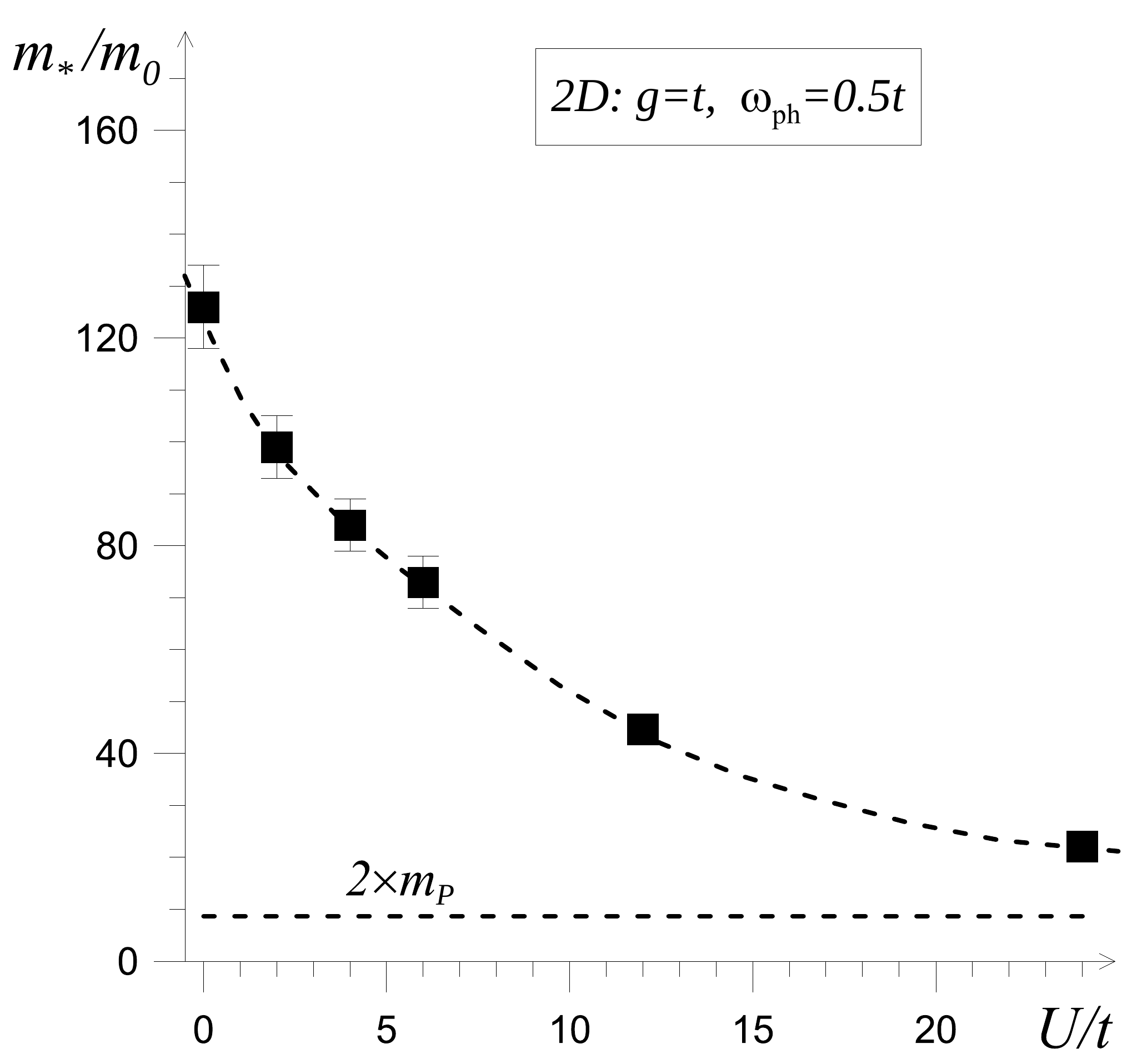}
\caption{Bipolaron energies (left) and effective masses (right) as functions of on-site
repulsion $U$ for coupling constant $g=t$ and adiabatic parameter $\gamma=1/16$.
Error bars for $E_{bi}$ are much smaller than symbol size.
}
\label{fig3}
\end{figure}
\begin{figure}[t]
\centering
\includegraphics[width=0.23\textwidth]{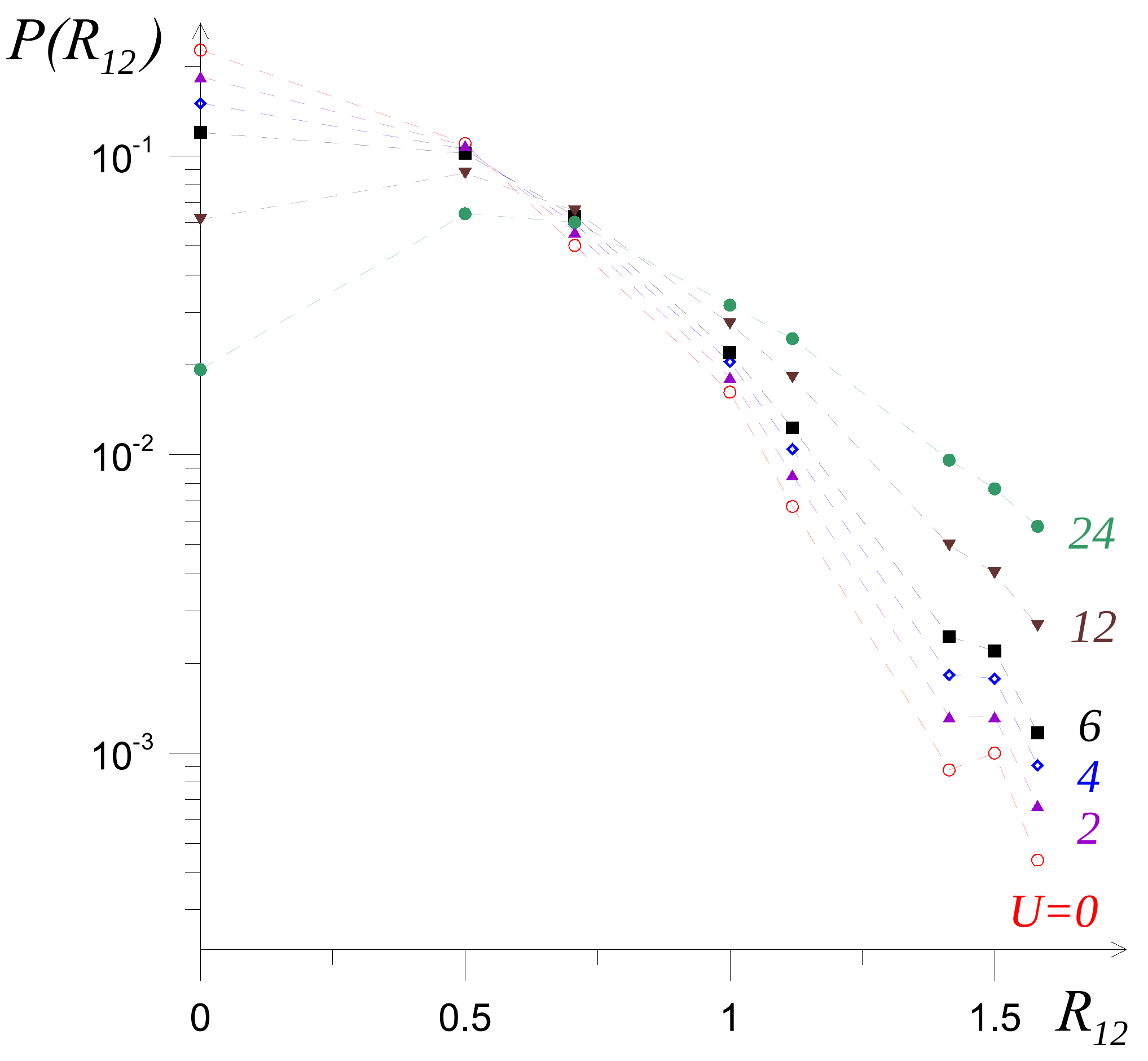}
\includegraphics[width=0.23\textwidth]{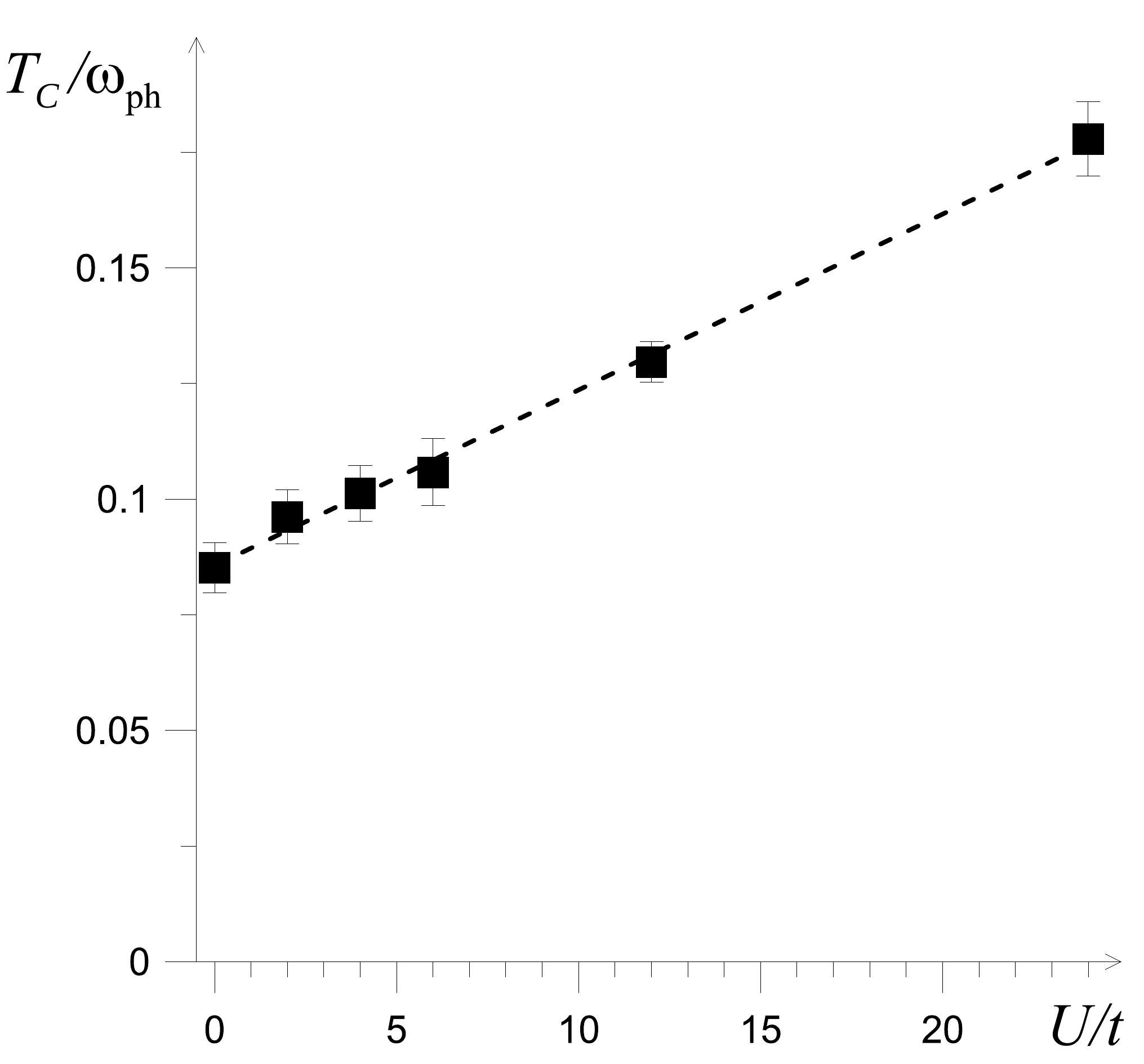}
\caption{Radial profiles for bipolaron states (left) and estimates of $T_c$ (right)
based on Eq.~(\ref{maxTc}) for the same set of parameters as in Fig.~\ref{fig3}.
}
\label{fig4}
\end{figure}

A pronounced maximum of $T_c$ at a finite $U$ does exist at a larger value of the coupling constant, as illustrated by the data for bipolaron states at $g=t$ and $\gamma=1/16$ in Figs.~\ref{fig3} and \ref{fig4}. Here the behavior of both the ground-state energy and the effective mass is qualitatively similar to what we had at $g=t/\sqrt{2}$ (see Fig.~\ref{fig1}). But the increase
of $R_*$ with $U$ is not as dramatic and does not overcompensate the decrease of $m_*$ even at $U\sim 20$.
The maximum of $T_c$ at $U > 24$ with the maximal value of $T_c$ significantly larger than at $U=0$ does exist
because for $U\to \infty $ the energy of the delocalized bipolaron state is extremely close to the bound state threshold.

{\it Conclusions}. Model (\ref{He})--(\ref{Hint}) of bond (bi)polarons (in any spatial dimensions) allows a controlled and efficient numeric solution by Monte Carlo methods formulated in the path-integral representation for the electron(s)
and either path-integral or real-space-diagrammatic representation for phonons.
This is also true for the generalizations of (\ref{He})--(\ref{Hint}) that include (i) dispersive bond phonons and (ii)
additional density-displacement couplings as in the Holstein model. For illustrative purposes, we presented simulations
of bipolaron states for model (\ref{He})--(\ref{Hint}) in two dimensions.

In the context of the bipolaron mechanism of high-temperature superconductivity, both the effective mass and the
size of bipolaron are equally important [see  Eq.~(\ref{maxTc})]. Our results show that the positive effect of having a
smaller effective mass can be readily overcompensated by the negative effect of having a larger bipolaron size; see Fig.~\ref{fig2}.

There is a range of parameters where the net effect of the strong repulsion between the electrons
is a substantial increase of the critical temperature for the superconducting transition; see Fig.~\ref{fig4}.
One may find this result rather counterintuitive given that normally the on-site repulsion suppresses Cooper pairing
in the $s$-channel.

{\it Acknowledgments.} NP and BS acknowledge support by the National Science Foundation
under Grant No. DMR-2032077.

\end{document}